\documentclass[aps,prb,reprint]{revtex4-1}
\usepackage{blindtext}
\usepackage{graphicx}
\usepackage{dcolumn}
\usepackage{bm} 
\usepackage{amssymb}
\usepackage{amsmath}
\usepackage{epsf}
\usepackage[utf8]{inputenc}
\usepackage{color}
\draft

\begin{document}

\preprint{APS/123-QED}

\title{Scaling behavior of Charge order melting in Magnetic field-Pressure-Temperature space of  2.5\% Al doped Pr$_{0.5}$Ca$_{0.5}$MnO$_3$ \\}

\author{Sudip Pal}
 
\author{Kranti Kumar}

\author{A. Banerjee}
\affiliation{%
 UGC DAE Consortium for Scientific Research\\
 Khandwa Road Indore 452001 
}%
%\date{\today}%
\begin{abstract}
Dc magnetic measurements across the charge ordering (CO) transition temperature (T$_{CO}$) in polycrystalline  Pr$_{0.5}$Ca$_{0.5}$Mn$_{0.975}$Al$_{0.025}$O$_3$ have been performed under simultaneous influence of external hydrostatic pressure (P) and magnetic field (H). We show the first experimental evidence that the melting of charge order instability obey an interesting scaling function, $\delta$T$_{CO}$/P$^\alpha$ = $f$(H/P$^\beta$) in H-P-T landscape, where $\delta$T$_{CO}$ is the suppression of T$_{CO}$ by P and H. Corresponding values of the exponents, $\alpha$ = 1.63 and $\beta$ = 0.33 have been extracted from data collapsing phenomena. Possible origin of such a scaling behavior has been discussed.
\end{abstract}
\maketitle

Careful experiments across phase transitions in variety of systems of interest to condensed matter physics have ignited flurry of theoretical activities to comprehend critical behavior which paved ways to outstanding concepts and models. Observation of data collapsing obeying scaling relationship falling within universality classes across phase transition for systems with diverse microscopic details is rather amazing {\color{blue}\cite{Stanley1971}}. Effect of relevant external parameters on the critical fluctuation, leading to renormalization of interactions, giving rise to scaling relations generating critical exponents according to the universality class for diverse systems and interactions is a success story of attempt to unify apparently incomprehensible many body systems. The triumph of this endeavour came from the studies of critical behaviour across continuous para- to ferromagnetic (FM) phase transition which is thoroughly worked out and verified in magnetization (M) - magnetic field (H) - temperature (T) landscape {\color{blue}\cite{SunilSir,SunilSirNL,AKP}}. Critical behavior has also been discussed in spin glasses, phase separated systems, dynamics of ferromagnets and high T$_C$ superconductors etc {\color{blue}\cite{Binderscaling, FMAvalanches, HTCSC, SunilSir2007}}. Aside these systems, such critical behavior has been surprisingly observed in many naturally occurring phenomena - viz., forest fire, earthquake, avalanches in granular media, or even in some of the functionalities happening within our brain {\color{blue}\cite{Forestfire,Cracklingnoise,Neural}}. Non-equilibrium systems may also show critical behavior, for example, thermal hysteresis across a first order phase transition (FOPT) shows a power law divergence with the temperature swipe rate {\color{blue}\cite{Debendetti,FOPTScaling, Hugues2005,KZScaling2013}}. Additionally, inclusion of random disorder can induce a continuous transition from a sharp first order transition in a rather pure system {\color{blue}\cite{Imry1979, Aizenman,Yun}}. Albeit, the critical behavior across such a disorder affected transition and the universality of such transition, if any, is still unsettled and not experimentally established in wide class of systems. In this context, for the first time it is shown here that the charge order (or CO) transition in a minimally doped (2.5\% Al at Mn site) prominent and robust CO manganese-oxide system (manganites), Pr$_{0.5}$Ca$_{0.5}$MnO$_3$ (PCMAO) also obeys an unexpected scaling relation in M, T, H, Pressure (P) landscape.

Manganites never ceased to puzzle the condensed matter scientists by its enriched complexity. Numerous works have been published on the CO transition in different mixed valence manganites {\color{blue}\cite{Moritomo,Roy,Tomioka,Dagotto}}. Though CO transition historically was first envisaged by E. Verway in Magnetite (Fe$_3$O$_4$)  in 1939 {\color{blue}\cite{Verway}},  its origin is being debated till recent times {\color{blue}\cite{Fe4O5,Baggari2017,Terletska}}. In mixed valence manganites when electrostatic coulomb repulsion dominates their kinetic energy, electrons localize at Mn sites as Mn$^{4+}$ and  Mn$^{3+}$ and the system becomes insulating. Moreover, electron localization suppresses double exchange (DE) interaction {\color{blue}\cite{DEAnderson}}, thereby pushing the system to the verge of superexchange (SE) mediated antiferromagnetic (AFM) state. Alternatively, a charge density wave scenario is also proposed for manganites {\color{blue}\cite{Littlewood2005, Baggari2017}}. Many other non-manganite families also show CO {\color{blue}\cite{Suard}} .

Effect of external P and H on the CO state in manganites can be explained using the semiclassical model of DE mechanism {\color{blue}\cite{Zener1951, DEAnderson, Millis1995}} as -

\begin{equation}
t_{eff}= t  \  Cos (\theta/2)
\end{equation}

Here ``$t_{eff}$" is the effective hopping integral of itinerant e$_g$ electrons. It is proportional to the bare hopping integral ($t$) and alignment of two adjacent t$_{2g}$ spin, $\theta$. In absence of spin scattering ($\theta$ = 0), the maximum possible effective hopping integral ($t_{eff}$) is equal to the bare hopping integral, $t$. Application of  external pressure squeezes the unit cell volume. Melting of a CO state by applied hydrostatic P is attributed to the straightening of  Mn$-$O$-$Mn bond angle and shrinkage of Mn$-$O bond length due to reduced distortion of MnO$_6$ octahedra by pressure {\color{blue}\cite{Moritomo}}. Such a modulation in bond angle and length increases the overlapping of outer ``d'' orbitals of a magnetic atom with the ``p'' orbital of intervening Oxygen atom and eventually modify ``t''. On the other hand, H always promotes electron itinerancy (``$t_{eff}$" in equ.1) by reducing  ``$\theta$'' between two spins. This reveals that the CO transition is influenced by charge, lattice and spins working in synergy, instead of  purely lying its origin on the charge carrier like in a canonical Mott-Hubbard transition {\color{blue}\cite{Mott1949}}.

We show here that the melting of CO state in PCMAO under simultaneous action of P and H obeys a scaling relation $\delta$T$_{CO}$ = P$^\alpha$ $f$(H/P$^\beta$), where $\delta$T$_{CO}$ is the suppression of T$_{CO}$ by P and H. It looks quite similar to the scaling relation M = $\epsilon$$^\beta$ $f$(H/$\epsilon$$^\beta$$^+$$^\gamma$) in M-H-T space around a continuous paramagnetic  to FM phase transition where $\epsilon$ [=(T-T$_C$)/T$_C$] is the reduced temperature and $\beta$,$\gamma$ describe the critical behavior of spontaneous magnetization and initial susceptibility respectively close to the critical temperature (or T$_C$). 

The details of sample preparation and characterization procedure are described elsewhere {\color{blue}\cite{SunilSirJPCM, S_Nair}}. Magnetization measurements are performed in $\pm$7 Tesla SQUID magnetometer (M/S Quantum design, USA). For external hydrostatic pressure, a Cu-Be pressure cell with pressure limit of 10 kbar (easyLabMcell 10) has been used. The reported pressure values are determined at low temperature by monitoring the superconducting transition temperature of Sn loaded inside the pressure cell {\color{blue}\cite{SDash}}. 

\begin{figure}[t]
\centering
\hspace{-0.6cm}
\includegraphics[scale=0.38]{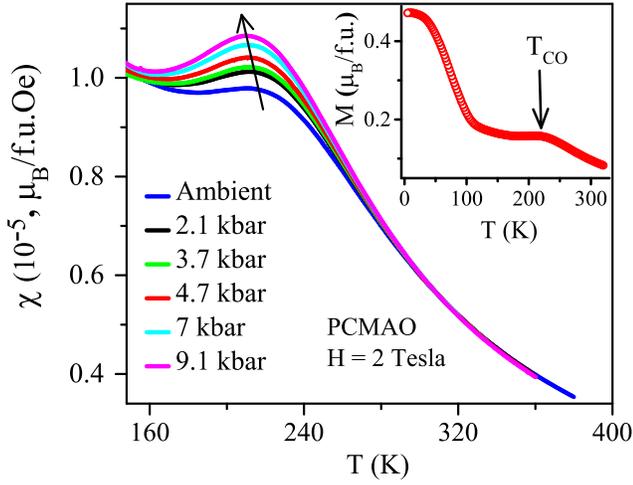}
\caption{\label{fig:epsart} Main panel shows the effect of pressure on the magnetic susceptibility around the charge ordering transition at H = 2 Tesla. Inset shows the entire M-T behavior of bulk PCMAO measured at H = 2 Tesla and ambient pressure.}
\end{figure}

\begin{figure}[t]
\centering
\hspace{-0.8 cm}
\includegraphics[scale=0.45]{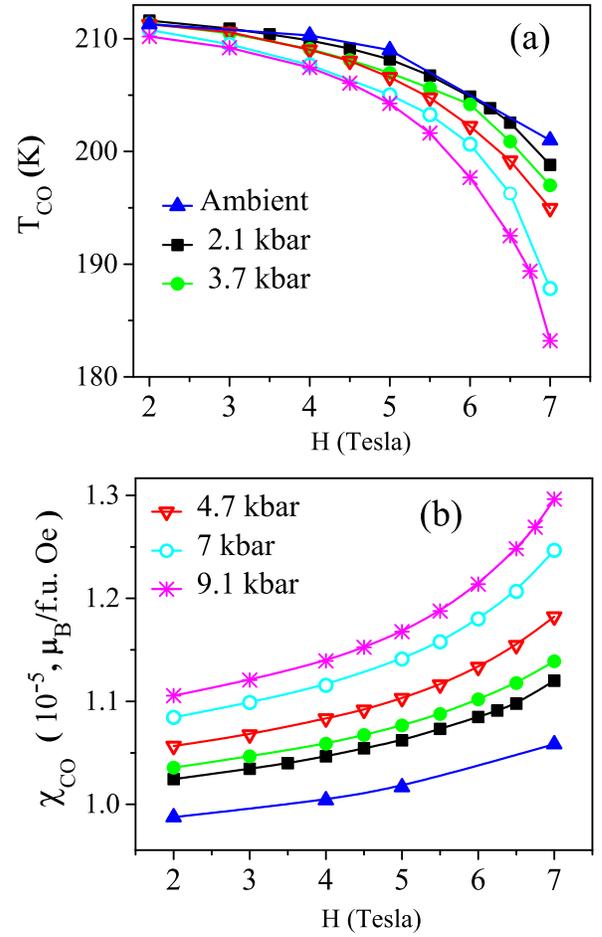}
\caption{\label{fig:epsart} Variation of  (a) T$_{CO}$ and (b) $\chi$ at T$_{CO}$ is plotted with H for different applied Pressures (Solid lines are guide to eye). Legends in both figure indicate the corresponding pressure values. }
\end{figure}

\begin{figure}[t]
\centering
\hspace{-0.6cm}
\includegraphics[scale=0.37]{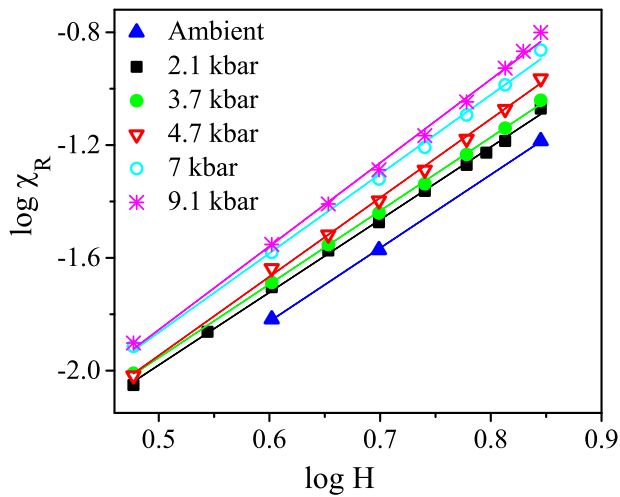}
\caption{\label{fig:epsart} Reduced susceptibility at T$_{CO}$ (see the text for details) which is derived from Fig. 2(b) is plotted with  H in log-log scale. Solid lines are linear fit to the data with slope varying between 2.6 to 2.9. }
\end{figure}

On cooling from paramagnetic state, PCMAO shows a prominent hump in magnetization around the charge ordering temperature, T$_{CO}$ = 211 K (see inset of Fig. {\color{blue}1}) and is subsequently followed by AFM and FM spin ordering respectively. Application of external P destabilizes the CO state and drives T$_{CO}$ progressively toward lower temperature (see Fig. {\color{blue}1}). Note that, at high T all susceptibility (M/H, where M is the dc magnetization) curves at different P are merged but as temperature is reduced toward the CO transition temperature, they start deviating from each other which is more pronounced just above CO. We have considered the temperature where susceptibility is maximum around CO as the T$_{CO}$. In Fig. {\color{blue} 2(a)} \& {\color{blue}(b)}, we have shown the variation of  T$_{CO}$ and susceptibility value at T$_{CO}$ with H at different values of P. Melting of the CO state by H is nonlinear and it melts more rapidly at higher P [See Fig. {\color{blue} 2(a)}]. Susceptibility also increases non-linearly with H [See Fig. {\color{blue} 2(b)}]. Interestingly, if we plot log-log plot of H versus reduced susceptibility at T$_{CO}$ (derived from Fig. {\color{blue}2(b)})  $\chi$$_R$  = $\chi$(H)- $\chi$(2 Tesla)/ $\chi$(2 Tesla) measured at different P, they varies linearly with slope varying between 2.6 to 2.9 [see Fig. {\color{blue}3}]. Here, $\chi$(2 Tesla) and $\chi$(H) are the susceptibility at 2 Tesla and other H measured at T$_{CO}$. It indicates that the susceptibility at T$_{CO}$ increases as a power law with field. On the other hand, suppression of CO temperature by field that we have quantified as $\delta$T$_{CO}$ = T$_{CO}$ (H) - T$_{CO}$ (2 Tesla), where T$_{CO}$ (2 Tesla) and T$_{CO}$ (H) are the CO temperature at 2 Tesla and other applied field, H respectively at different P, increases as P increases (not shown here). Here we have subtracted the  T$_{CO}$ values at all P and H by the T$_{CO}$ measured at our lowest measuring field of 2 Tesla and the corresponding P and designated it as $\delta$T$_{CO}$. In the quest to see the mutual influence of P and H on the suppression of T$_{CO}$, we started with plotting $\delta$T$_{CO}$/P versus H/P. Then we varied the powers of P in the Y and X-axis through an iterative process to get a good data collapse similar to the procedure followed in Ref. [4] and found that rescaling the Y and X-axis  as $\delta$T$_{CO}$/P$^\alpha$ and H/P$^\beta$, respectively, with $\alpha$= 1.63 and $\beta$= 0.33, collapse all the $\delta$T$_{CO}$ vs H curves at different P onto a single curve as shown in the main panel of Fig. {\color{blue}4} and this curve merges with the $\delta$T$_{CO}$ vs H data measured at ambient pressure (See the upper inset of Fig. {\color{blue}4}). In this inset, we have shown the $\delta$ T$_{CO}$ versus H data at ambient pressure and the rescaled data only for P=2.1 kbar for the sake of clarity. In the lower inset, we have shown the double log plot of the data shown in the main panel of Fig. {\color{blue}4}. It shows considerable data collapse and varies linearly. Such data collapse signifies that the melting of CO state in simultaneous action of P and H obeys a single scaling relation: 

\begin{equation}
\delta T_{CO}/P^\alpha = f(H/P^\beta). 
\end{equation}

\begin{figure}[t]
\centering
\hspace{-0.5cm}
\includegraphics[scale=0.3]{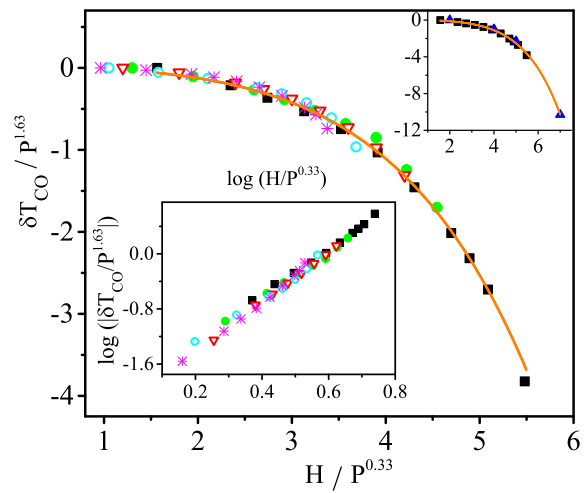}
\caption{\label{fig:epsart} Main panel shows the collapsing of T$_{CO}$ values in H-P-T space after rescaling the X and Y axis of Fig. 2(a). Lower inset shows the log-log plot of the data shown in the main panel. Upper inset shows the collapse of $\delta$T$_{CO}$ versus H data at ambient pressure with the rescaled data at P=2.1 kbar. Symbols present the pressure values as in Fig. 3.}
\end{figure}

We note here that, the indices obtained from data collapsing [Fig. {\color{blue}4}] are surprisingly not arbitrary, because, a power law dependence of physical properties with critical index of classic value 1/3 is very familiar for three dimensional (3D) systems outside the regime of mean field theory and 1.63 is also very close to the value of ($\beta$$+$$\gamma$) found around a critical region belonging to the 3D Ising or Heisenberg universality class of a continuous transition. The system studied here is not a mean field system in the probing temperature and field range because it violates the basic assumption of mean field theory of absence of fluctuations. $\chi$$^{-1}$ vs T (not shown here) at H= 0.01 Tesla and ambient P fits well with the Curie-Weiss law only at high T ($>$ 320 K) with a curie constant of 212 K and $\mu$$_{eff}$ = 4.83 $\mu$$_B$/f.u. which is larger than the expected spin only value of 4.36  $\mu$$_B$/f.u. Moreover, the inverse susceptibility deviates upward from Curie-Weiss behavior much above the CO transition. These features are reminiscent of  predominant short range FM correlation/fluctuations at high temperature {\color{blue}\cite{Mydosh}}. W. Bao et al have studied this kind of correlation above CO in Bi$_{1-x}$Ca$_x$MnO$_3$ (0.74 $\leqslant$ x $\leqslant$ 0.82) by neutron scattering and bulk magnetization measurements {\color{blue}\cite{Bao}}. Later on, similar spin correlation was reported in Pr$_{1-x}$Ca$_x$MnO$_3$ (x$\leqslant$ 0.50)  prior to the CO transition.  Below CO, these correlations are replaced by AFM correlations due to carrier localization and the magnetic susceptibility decreases. This is considered to be the reason for the prominent hump observed in magnetization around the CO transition as also observed in Fig.{\color{blue}1} {\color{blue}\cite{Kajimoto1998, Aladene2006}}. However, CO transition is generally considered as a first order phase transition. Thus the observation of scaling behavior here is not trivial. 
Notwithstanding the above, we can write an expression for T$_{CO}$ from equ. 1, as -

\begin{equation}
\textrm{T}_{CO} = \phi (\textrm{P}) * g(\textrm{H})
\end{equation}

where $\phi$(P) is a function of P which depends on the change in $t$ and electron-phonon coupling strength by the applied P and $g$(H) is a function of H which depends on ``$\theta$". The exact functional dependence will of course depend on the intrinsic details of magneto-electronic interactions of the sample. So following equ. 3, we can claim that it is possible to rescale  T$_{CO}$ by some function of P such that it follows a scaling function, $\delta$T$_{CO}$/$\phi$(P) = $g$(H). But here we have few important observations: (1) reduced susceptibility at T$_{CO}$ varies as power law with H [Fig. {\color{blue}3}], (2) we need to additionally rescale H to merge T$_{CO}$ values, and (3) which is significant, both functional dependence are in the form of power law with indices close to 3D Ising and Heisenberg universality class [Fig. {\color{blue}3}]. Therefore, data collapsing observed here cannot be construed as merely fortuitous or mathematical. Such data collapsing hints toward a scaling behavior of the charge solid to liquid melting process.
In this context, it is noteworthy that the charge order transition observed in manganites has been generally described as a first order transition because of finite latent heat observed in specific heat and structural changes across the transition {\color{blue}\cite{AKR2001, Dagotto}}. Moreover, in this case the parent compound Pr$_{0.5}$Ca$_{0.5}$MnO$_3$ (PCMO) is a robust charge ordered insulating system and the temperature induced CO transition itself is argued to be a first order {\color{blue}\cite{Moritomo}}. 

In PCMAO random substitution of 2.5\% non-magnetic Al in Mn site has drastic impact on the magnetic properties. For example, (i) charge order transition is weakened; it is shifted to lower temperature with increasing amount of Al substitution. Though some anomaly is observed in magnetic entropy, no indication of latent heat could be inferred from specific heat measurement {\color{blue}\cite{SunilSirJPCM, AB2009}}. (ii) Significantly, the robust AFM-insulating ground state of the parent system has surprisingly changed to FM-metallic state {\color{blue}\cite{AB2006, Dagatto2009}}, while substitution of Al neither affect the lattice structure significantly, nor contribute to any magnetic interaction. Substitution of Al created frozen random disorder in CO lattice which would broaden the first-order transition {\color{blue}\cite{Imry1979}} and possibly drive the first order CO transition in parent compound into a continuous transition. In that case, a scaling behavior can appear {\color{blue}\cite{Imry1979, Aizenman,Yun}}. However, it throw up a challenge to find the exact underlying mechanism behind such mutual scaling behavior of three thermodynamic parameters i.e. the CO transition temperature, magnetic field and  hydrostatic pressure within such a wide range in PCMAO. 
 
 Further,  in a first order transition, two or more phases co-exist at phase boundary and none of these phases is critical as in a continuous phase transition. However, the phase coexistence of two or more phases across a first order transition can also be studied using renormalization group fixed point approach.  Here, the order parameter shows critical behavior with the critical exponent equals to zero and the ``coherence or persistence length'' relating to the occurrence of co-existing long range order phases diverges with an exponent of 1/3 for a system with spatial dimension of three {\color{blue}\cite{Nienhuis1975,Fisher1982,Stefano2018}}. Recently, intensive works are directed toward this field {\color{blue}\cite{Liang2017, Vicari2017}}. In this context, the random field Ising model (RFIM) has been extensively investigated for nearly three decades, where, a FOPT is predicted in both extensive Monte-Carlo simulations and experiments with few observables showing critical (``quasicritical'') behavior {\color{blue}\cite{Young1985, Birgeneau1985}}. Although, the order of phase transition in RFIM is still unresolved, a recent rigorous numerical simulation has found critical behavior across a weak FOPT {\color{blue}\cite{Maiorano2007}}. A weak first order Mott transition also shows critical behavior {\color{blue}\cite{Furukawa2018}}.

To summarize, we have studied the melting of the CO state in 2.5\% Al doped Pr$_{0.5}$Ca$_{0.5}$MnO$_{3}$  under simultaneous application of external hydrostatic pressure and magnetic field through the magnetization measurement. We have shown that this melting process follows an interesting scaling relation in H-P-T space. The exponents obtained from collapsed data from a rigorous process are close to the 3D Ising and Heisenberg universality class. Such scaling behavior can arise in a disorder broadened first order transition culminating to a continuous transition. Our study brings out an important query about the nature of CO transition through the observed critical behavior in the melting of CO by T, P and H. It also instigates bringing in other relevant parameters into the study of CO state like Electric Field.  Moreover, the present study not only calls for serious scrutiny of the CO state under simultaneous effect of T, P and H in other systems but also expect to stimulate theoretical studies around the critical regime of such disorder broadened transitions. More importantly, this study brings out the importance of more than two thermodynamic parameters to properly fix or designate the critical end point for a transition belonging to known universality class.  

We would like to convey our gratitude to Dr. S. B. Roy  for many relevant discussions.

\end{document}